\documentclass[12pt]{article}
\usepackage[utf8]{inputenc}
\usepackage{graphicx}
\usepackage{xcolor}
\usepackage{natbib}
\usepackage[toc,page]{appendix}
\usepackage{geometry}
\usepackage{microtype}
\usepackage{setspace}
\usepackage{fullpage}
\usepackage{tabularx}
\usepackage{hyperref}% 

\usepackage{amsmath, amssymb, amsthm}
\theoremstyle{plain}

\newtheorem{theorem}{Theorem}

\theoremstyle{definition}

\theoremstyle{remark}

\DeclareMathOperator{\diag}{diag}
\DeclareMathOperator*{\argmax}{arg\,max}

\newcommand{\E}{\mathsf{E}}
\newcommand{\I}{\mathbb{I}}

\newcommand{\cov}{\mathrm{cov}}

\newcommand{\pr}{\mathrm{P}}

\DeclareMathOperator{\bdiag}{bdiag}

\newcommand{\tr}{\operatorname{tr}}
\newcommand{\tsp}{\mathrm{\scriptscriptstyle T}}
\newcommand{\rN}{\mathrm{N}}
\newcommand{\rU}{\mathrm{U}}

\newcommand{\R}[1]{\mathbb{R}^{#1}}

\def\T{{ \mathrm{\scriptscriptstyle T} }}
\newcommand{\Llh}{\mathcal{L}}

\title{Universal inference for variance components}
\author{Yiqiao Zhang$^\star$ \quad Karl Oskar Ekvall$^{\star, \dagger}$ \quad
Aaron J. Molstad$^{ \ddagger}$\\
{\normalsize $^\star$Department of Statistics, University of Florida} \\
{\normalsize $^\dagger$Division of Biostatistics, Institute of Environmental
Medicine, Karolinska Institutet}\\
{\normalsize $^\ddagger$School of Statistics, University of Minnesota}\\
{\tt \normalsize yiqiaozhang@ufl.edu \quad k.ekvall@ufl.edu \quad
amolstad@umn.edu}}
\date{\normalsize \today}
\date{\today} 

\begin{document}

\maketitle

\begin{abstract}
    We consider universal inference in variance components models, focusing on settings where the parameter is near or at the boundary of the parameter set. Two cases, which are not handled by existing state-of-the-art methods, are of particular interest: (i) inference on a variance component when other variance components are near or at the boundary, and (ii) inference on near-unity proportions of variability, that is, one variance component divided by the sum of all variance components. Case (i) is relevant, for example, for the construction of componentwise confidence intervals, as often used by practitioners. Case (ii) is particularly relevant when making inferences about heritability in modern genetics. For both cases, we show how to construct confidence intervals that are uniformly valid in finite samples. We propose algorithms which, by exploiting the structure of variance components models, lead to substantially faster computing than naive implementations of universal inference. The usefulness of the proposed methods is illustrated by simulations and a data example with crossed random effects, which are known to be complicated for conventional inference procedures.
\end{abstract}

\onehalfspacing

\section{Introduction}

Variance components models are used routinely in a wide variety of scientific applications. Often times, multiple sources of variation are present, in which case practitioners want to understand the degree of total variation attributable to each source. In epidemiology, for example, researchers want to understand how a trait's variability is affected by additive genetic effects and the environment \citep{heckerman2016linear}, or by additive genetic effects and gene-environment interactions \citep{pazokitoroudi2024scalable}. Similarly, in statistical genetics, genetic effects can be partitioned by chromosome \citep{yang2011genome}, partitioned into purely additive effects versus genetic interaction effects \citep{bloom2015genetic,vitezica2013additive}, partitioned into additive and dominance effects of SNP markers \citep{da2014mixed}, among many other partitioning approaches \citep[e.g., see][]{runcie2019fast}. 

More generally, hierarchical and multilevel mixed models aim to quantify the degree of a random variable's variability that can be attributed to distinct sources of variation \citep{goldstein2011multilevel,kreft1998introducing, lee1996hierarchical, rasbash1994efficient}. For example, the variability in student test scores may be attributed to variation arising from classroom effects, school-level effects, or community-wide effects. 

One widely used model for capturing multiple sources of variation is the variance components model. 
A variance components model assumes a vector $Y \in \R{n}$ satisfies, for some known symmetric and positive semi-definite $K_1, \dots, K_M$, $M$ a positive integer,
\begin{equation}\label{eq:mult_varcomp}
    Y \sim \rN(0, \sigma^2_{K_1} +\cdots+ \sigma^2_MK_M + \sigma^2_{M+1}I_n)   ,
\end{equation}
where $\sigma^2_1, \dots, \sigma^2_M$ are the variance components. Here, we assume $\E(Y) = 0$ for simplicity but later allow $\E(Y) = X\beta$ for known $X \in \R{n\times p}$ and unknown $\beta \in \R{p}$. The distribution in
\eqref{eq:mult_varcomp} sometimes results from a random effects model:
\[
    Y = Z_1 U_1 + \cdots + Z_M U_M + E,
\]
where, independently for each $m \in \{1, \dots, M\}$, random effects satisfy $U_m \sim \rN(0, \sigma^2_m I_{q_m})$ for some $q_m \in \{1, \dots, m\}$, and $Z_m \in \R{n\times q_m}$. The error term $E \sim \rN(0, \sigma^2_{M + 1}I_n)$ is independent of the random effects. Then, for every $m \leq M$, $K_m = Z_mZ_m^\T$, so that the rank of $K_m$ is at most $q_m \leq n$. To avoid degenerate distributions, we will typically assume the error variance is nonnegative, $\sigma^2_{M + 1} > 0$. 

Our focus is hypothesis tests, or inferences more generally, for the variance components and the proportions 
\begin{equation}
    h^2_m = \frac{\sigma^2_m}{\sum_{m=1}^{M+1}\sigma_m^2}, \quad m\in\{1,\dots,M\}.
\end{equation}
The parameter $h_m^2$ is often interpreted as the proportion of variability attributable to sources encoded by $K_m$, $m \in \{1, \dots, M\}$. In statistical genetics, for example, $K_m$ can encode the genetic similarly of individuals' $m$th chromosome. Then, $h^2_m$ is the proportion of variance in the outcome explained by the $m$th chromosome's SNP genotypes \citep{yang2011genome}. To facilitate inference on the $h^2_m$ we consider a reparameterization of \eqref{eq:mult_varcomp} in terms of $\theta = (h^2_1, \dots, h^2_M, \tau^2)^{\T}$, where $\tau^2 = \sum_{m = 1}^{M + 1}\sigma^2_{m}$. Then, the parameter set $\Theta \subseteq \R{M + 1}$ is the set of $\theta$ such that $\sum_{m = 1}^M h_m^2 < 1$, $h^2_m \geq 0$ for all $m$, and $\tau^2 > 0$. With these parameters $Y$ is multivariate normal with mean zero and covariance matrix
\begin{equation} \label{eq:varcomp_sigma}
   \Sigma = \Sigma(\theta) = \tau^2\left\{h_1^2 K_1 + \cdots + h_M^2 K_M + \left(1 - \sum_{m = 1}^M h_m^2\right)I_n\right\} \in \R{n\times n}.
\end{equation}

\begin{figure}
    \centering
    \includegraphics[width=1\linewidth]{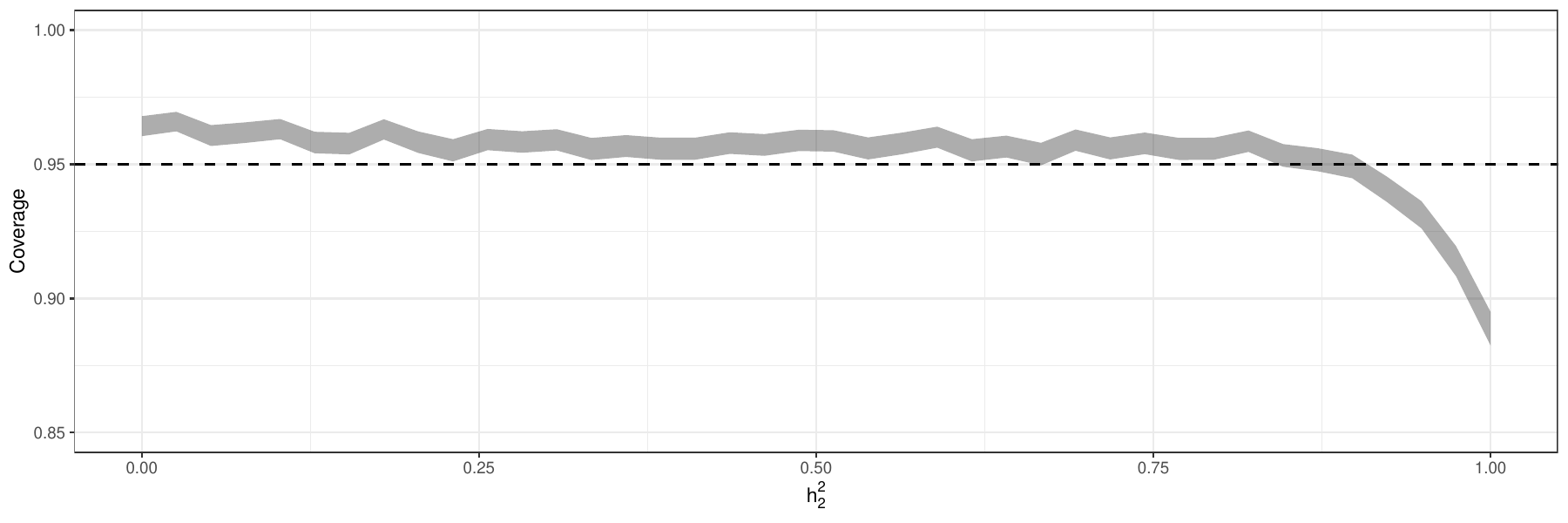}
    \caption{Coverage probabilities for a score-based confidence interval for $h_1^2 = 0$, for different values of the nuisance parameter $h_2^2$. Estimates based on 10,000 replications. The shaded region indicates 95\% confidence bands.}
    
    \label{fig:motivationforchp3}
\end{figure}

Inference on the $h^2_m$ is complicated in general because one or more of them are often close to zero. That is, the true parameter is often at or near the boundary of the parameter set. When $M = 1$, so that there is a single $h^2_m$, there are methods based on inverting score test-statistics \citep{zhang_fast_2025} and simulation-based methods \citep{Crainiceanu.Ruppert2004,schweiger2018using, Schweigeretal2016}. However, the supporting theory for the simulation-based methods is not applicable when $M > 1$, and the score-based confidence intervals have non-nominal coverage probability when there are nuisance parameters near the boundary. For example, a score-based confidence interval for $h^2_1$ often has non-nominal coverage probability if another $h^2_m$, $m \neq 1 $, is close to zero or one. Figure \ref{fig:motivationforchp3} illustrates this issue in a setting where $M = 2$; the parameter of interest is $h_1^2$, whose true value is zero; and $h_2^2$ is a nuisance parameter whose true value is on the horizontal axis. Clearly, the coverage probability for the confidence interval for $h_1^2$ is affected by the true value of the nuisance parameter, especially when it is near one. Somewhat informally, the issue is that test-statistics for $h_1^2$ depend on a constrained maximum likelihood estimator of $h_2^2$, and that estimator behaves irregularly near the boundary. 

The issues are easier to deal with when there are no nuisance parameters as, then, score-based test-statistics can be evaluated at the null hypothesis parameter vector, no estimation needed \citep{zhang_fast_2025,ekvall2025uniform}. However, even without nuisance parameters, state-of-the art methods are unreliable when a $h_m^2$ is near unity. The main reason is that points where $\sum_{m = 1}^M h_m^2 = 1$ are often hard boundary points, in the sense that the likelihood cannot be extended to such points, while points where a  $h^2_m = 0$ are soft boundary points \citep{elkantassi2023improved}.

Here, we consider methods for componentwise inference for variance components in the presence of nuisance parameters. The methods are based on universal inference, and in particular split likelihood ratio tests \citep{wasserman2020universal}. Consequently, the proposed tests and confidence intervals are uniformly valid in finite samples, regardless of how close to a soft or hard boundary the parameter is. Thus, even when there are no nuisance parameters, the methods studied here can be preferable to common ones, which are often motivated by asymptotic theory. We describe universal inference and its application to our setting in more detail in Section \ref{sec:randomsplitLRT}. Briefly, the method relies on data splitting, where one set of data is used to estimate parameters and the other to carry out the test. In our setting, where there is a potentially complicated dependence structure, more care is needed when randomly splitting the data than in settings with independent and identically distributed observations. A main concern with universal inference is a lack of power compared to classical parametric methods, which we address by using a randomized version of the split likelihood ratio test \citep{ramdas2023randomized}.

There are several computational challenges with applying universal inference to \eqref{eq:mult_varcomp}, especially when $n$ is large. To address such issues we develop efficient algorithms for several special cases of interest. For example, in a setting with crossed random effects, which are known to complicate both computation and theory, we decrease the required time by several orders of magnitude compared to a naive implementation.

% \section{Model}\label{sec:VarianceCompsModel}
% %\subsection{Variance Components Models} \label{sec:VarianceCompModels}
% Consider a multiple variance components model 
% \begin{equation}
%     Y \sim \rN(0, \sigma^2_1K_1 +\cdots+ \sigma^2_MK_M + \sigma^2_{M+1}I_n)  
% \end{equation}
% is of interest,
% where $\{K_m\}_{m=1}^M$ are known covariance matrices. Suppose $\sigma^2 = \sum_{m=1}^{M+1}\sigma^2_m$ and $h^2_m = \sigma^2_m/\sigma^2$, $ m\in\{1, 2, \dots, M\}$. Thus, let $\theta = (h_1^2, h_2^2, \sigma^2)^{\T}$ denote the vector of unknown parameters and let $ \Theta= [0,1) \times [0,1) \times (0, \infty)$ denote the parameter set. We propose a universal method with several computational improvements for special cases in the following sections to inference on parameters of interest. %Thus variance of Y, $\Sigma=\sigma^2(h_1^2K_1 + h_2^2K_2 + (1-h_1^2-h_2^2)I_n)$. 

\section{Randomized Split Likelihood Ratio Test}\label{sec:randomsplitLRT}
Let $Y \in \R{n}$ be a random vector with density $f_{\theta^*}(y)$, $\theta^* \in \Theta$ being the true value of the parameter. Let $\Llh_Y(\theta) = f_\theta(Y)$ be the (random) likelihood function; that is, the density of $Y$, evaluated at $Y$, considered as a function of $\theta$ with domain $\Theta$. In our setting the density and likelihood correspond to \eqref{eq:mult_varcomp}, but universal inference applies more generally. Suppose we wish to test test $H_0: \theta^* \in \Theta_0 \subseteq \Theta$ versus $H_A: \theta^* \in \Theta\setminus \Theta_0$. Recall that a conventional likelihood ratio test (LRT) can be based on the statistic
\[
    \frac{\sup_{\theta \in \Theta}\Llh_{Y}(\theta)}{\sup_{\theta \in \Theta_0}\Llh_{Y}(\theta)}.
\]
Under well-known regularity conditions, two times the logarithm of this statistic has an asymptotic chi-square distribution, which can be used to construct tests with asymptotically correct size. However, such regularity conditions do not hold in our setting, so we instead consider the split LRT.

Let $Y_{(0)} \in \R{n_{(0)}}$ and $Y_{(1)} \in \R{n_{(1)}}$ be a random partition of $Y$, where $n_{(0)} + n_{(1)} = n$. That is, every element of $Y$ is in exactly one of the $Y_{(i)}, i \in \{0, 1\}$. The randomization is done independently of $Y$, but it need not be uniform on the set of possible partitions of given sizes. Let $\Llh_{Y_{(1)}}(\theta)$ be the likelihood based on $Y_{(1)}$ and  $\Llh_{Y_{(0)} | Y_{(1)}}(\theta)$ likelihood based on the conditional distribution of $Y_{(0)} $ given $ Y_{(1)}$. Let also $\hat{\theta}_1\in\argmax_{\theta \in \Theta}\Llh_{Y_{(1)}}(\theta)$ and $\hat{\theta}_0\in\argmax_{\theta \in \Theta_0}\Llh_{Y_{(0)} | Y_{(1)}}(\theta)$, assuming they exist. Then the split likelihood-ratio statistic is 
\begin{equation}\label{eq:test_stat_slrt}
    T_n = \frac{\Llh_{Y_{(0)} | Y_{(1)}}(\hat{\theta}_1)}{\Llh_{Y_{(0)} | Y_{(1)}}(\hat{\theta}_0)}.
\end{equation}
The test that rejects $H_0$ if $T_n > 1/\alpha$ is valid at level $\alpha \in (0, 1)$ \citep{wasserman2020universal}; that is, the size of the test is at most $\alpha$, and hence confidence regions obtained by inverting the test have coverage probability at least $1 - \alpha$. In fact, the test remains valid, and has greater power, if rejection is instead based on comparing to a uniform random variable \citep{ramdas2023randomized}. We state these facts formally in the following known result and provide a proof in the Appendix for completeness. It will be important for later to note that the same result and proof holds if, in \eqref{eq:test_stat_slrt}, $\hat{\theta}_1$ is replaced by any other estimator based on $Y_{(1)}$ only. Similarly, $\hat{\theta}_0$ can be replaced by any $\tilde{\theta}$ such that $\Llh_{Y_{(0)}\mid Y_{(1)}}(\tilde{\theta}) \geq \Llh_{Y_{(0)}\mid Y_{(1)}}(\hat{\theta}_0)$.

\begin{theorem}\label{thm:validTest}
    The split likelihood ratio test that rejects if $T_n > 1 / \alpha$ is a uniformly valid level $\alpha$ test, i.e., for any $\Theta_0 \subseteq \Theta$ and $\theta^* \in \Theta_0$, it holds that $\pr_{\theta^*}(T_n> 1/\alpha) \leq \alpha$. Moreover, the randomized split likelihood ratio test that rejects if $T_n > U/\alpha$, with $U\sim \rU(0, 1)$ independently of $T_n$, is also uniformly valid, and it is more powerful than the split likelihood ratio test.
\end{theorem}

% \subsection{Application to generalized linear mixed models}

% Suppose that the elements of $Y \in \R{n}$ are conditionally independent given
% $U \in \R{q}$, with
% \[
%     f_{\theta}(y\mid u) = \prod_{i=1}^n f_{\theta}(y_i\mid u) = \prod_{i=1}^n \exp\{y_i \eta_i - c(\eta_i) \} = \exp\left[\sum_{i=1}^n \{y_i \eta_i - c(\eta_i)\}\right],
% \]
% where $\eta_i = X_i^\tsp \beta + Z_i^\tsp U$. Here, $c$ is a cumulant function satisfying $c'(\eta_i) = \E(Y_i \mid U)$ and $c''(\eta_i) = \var(Y_i \mid U)$. Terms not depending on $\theta$ or $U$ are absorbed in the density's dominating measure. The likelihood is therefore
% \[
%     \Llh(\theta) = \int f_{\theta}(Y\mid u)f_\theta(u)du = \int \exp\{-g(u, \theta)\}du,
% \]
% where
% \[
%     g(u, \theta) = g(u, \theta; Y, X, Z) = -\sum_{i = 1}^n \{Y_i \eta_i - c(\eta_i)\} - \log f_\theta(u).
% \]
% With $\tilde{u} = \tilde{u}(\theta) \in \argmin g(u, \theta)$ and $H(\theta) = \nabla_u ^2 g\{\tilde{u}(\theta), \theta\}$, the Laplace approximation is
% \[
%     \tilde{\Llh}(\theta) = \vert H(\theta)\vert^{-1/2} \exp[-g\{\tilde{u}(\theta), \theta\}],
% \]

\subsection{Application to Variance Components} \label{sec:app_varcomp}

Assume \eqref{eq:mult_varcomp} with the parameterization in \eqref{eq:varcomp_sigma} and let
\[
    \Psi(h^2) = \sum_{m = 1}^M h_m^2 K_m + \left(1 - \sum_{m = 1}^M h_m^2 \right)I_n, ~~h^2 = (h_1^2, \dots, h_M^2)^\T.
 \]
 Then, ignoring additive terms not depending on $\theta$,
\begin{align*}
    l_Y(\theta) &= \log \Llh_Y(\theta) = -\frac{1}{2}\left(\log \vert \Sigma(\theta)\vert + Y^\T \Sigma(\theta)^{-1}Y\right) \\
    &= -\frac{1}{2}\left\{n\log(\tau^2) + \log \left\vert \Psi(h^2)\right\vert + \frac{1}{\tau^2}Y^\T \Psi(h^2)^{-1}Y\right\}.
\end{align*}
Because the partitioning is independent of $Y$, $Y_{(0)}$ and $Y_{(1)}$ are jointly multivariate normal. Thus, the marginal log-likelihood for $Y_{(1)}$ is similar to that for $Y$, with $Y_{(1)}$ and $\Sigma_{(11)}$ in place of $Y$ and $\Sigma$, respectively.
Also,
\[
    Y_{(0)}\mid Y_{(1)} \sim  \rN\left(\Sigma_{(01)}\Sigma_{(11)}^{-1}Y_{(1)}, \Sigma_{(00)}-\Sigma_{(01)}\Sigma_{(11)}^{-1}\Sigma_{(10)}\right),
\]
where $\Sigma_{(ij)} = \Sigma_{(ij)}(\theta) = \E_{\theta}(Y_{(i)}Y_{(j)}^\T) -  \E_{\theta}(Y_{(i)}) \E_{\theta}(Y_{(j)})^\T$, $i, j \in \{0, 1\}$. 
Thus, the conditional log-likelihood $  l_{Y_{(0)}\mid Y_{(1)}}(\theta)$  is
\begin{align} \label{eq:lY0|1_whole}
   &=  -\frac{1}{2} \log|\Sigma_{(00)}-\Sigma_{(01)}\Sigma_{(11)}^{-1}\Sigma_{(10)}| \\
    &\quad -\frac{1}{2}\left\{(Y_{(0)}-\Sigma_{(01)}\Sigma_{(11)}^{-1}Y_{(1)})^\T(\Sigma_{(00)}-\Sigma_{(01)}\Sigma_{(11)}^{-1}\Sigma_{(10)})^{-1} \right. \notag \\ 
    &\quad \quad\left. (Y_{(0)}-\Sigma_{(01)}\Sigma_{(11)}^{-1}Y_{(1)})\right\}.\notag
\end{align}
 We can write the $\Sigma_{(ij)}$ in terms of the $K_m$ as follows. For each $m \in \{1, \dots, M\}$, let $K^m_{(ij)}$ be the matrix obtained from $K_m$ by keeping only the rows with indices corresponding to observations in $Y_{(i)}$ and columns with indices corresponding to observations in $Y_{(j)}$. Then
\[
    \Sigma_{(ij)} = \sum_{m = 1}^M h_m^2 K^m_{(ij)} +  \I(i = j) \left( 1 - \sum_{m = 1}^M h_m^2\right) I_{n_{(0)}}, ~~ i, j \in \{0, 1\},
\]
where $\I(\cdot)$ is an indicator function and $n_{(0)}$ the number of elements in $Y_{(0)}$.

Several challenges with implementing universal inference are evident from \eqref{eq:lY0|1_whole}. In general, finding the maximizers $\hat{\theta}_0$ and $\hat{\theta}_1$ is nontrivial and requires numerical optimization. To find $\hat{\theta}_1$, we use off-the-shelf, gradient-based methods to maximize $l_{Y_{(1)}}(\theta)$. To that end, note that for any element of $\theta$, say $\theta_j$,
\begin{equation} \label{eq:sigma_derivs}
    \frac{\partial l_{Y}(\theta)}{\partial \theta_j} = -\frac{1}{2}\tr\left[\left\{\Sigma(\theta)^{-1} - \Sigma(\theta)^{-1}YY^\T \Sigma(\theta)^{-1}\right\} \frac{\partial \Sigma(\theta)}{\partial \theta_j}\right],
\end{equation}
and similarly for $l_{Y_{(1)}}(\theta)$ but with $Y_{(1)}$ and $\Sigma_{(11)}$ in place of $Y$ and $\Sigma$, respectively. When $j = M + 1$, so that $\theta_j = \tau^2$, \eqref{eq:sigma_derivs} is
\begin{align*}
    &-\frac{1}{2}\tr\left[\Sigma(\theta)^{-1}\left\{I_n - \tau^{-2} YY^\T \Psi(h)^{-1}\right\}\Psi(h^2)\right]\\
    &= -\frac{1}{2}\tr\left[\tau^{-2}\left\{I_n - \tau^{-2} YY^\T \Psi(h)^{-1}\right\}\right],
\end{align*}
which vanishes if evaluated at $\tau^2 = n^{-1} Y^\T \Psi^{-1}(h^2)Y$. It is routine to show this stationary point is in fact a global partial maximizer. Thus, an algorithm for finding $\hat{\theta}_1$ can alternate between updating optimization variables corresponding to $\tau^2$ and $h^2$, with the former update being available in closed form. By contrast, first order conditions for $h^2$ based on \eqref{eq:sigma_derivs} cannot in general be solved analytically. Thus, we update $h^2$ using a gradient-based step. Equivalently, we use gradient-based methods to maximize the profile log-likelihood $h^2 \mapsto l_{Y_{(1)}}\{h^2, \tilde{\tau}^2_{(1)}(h^2)\}$, where
\[
    \tilde{\tau}_{(1)}^2(h^2) = n_{(1)}^{-1} Y_{(1)}^\T \Psi_{(11)}^{-1}(h^2)Y_{(1)}.
\]
The derivatives of the profile log-likelihood, $\partial l_{Y_{(1)}} \{h^2, \tilde{\tau}^2(h)\} / \partial h_m^2$, are
\begin{align*}
  & \left. \frac{\partial l_{Y_{(1)}}(h^2,\tau^2)}{\partial h_m^2}\right\vert_{h^2, \tilde{\tau}_{(1)}^2(h^2)} +  \left. \frac{\partial l_{Y_{(1)}}(h^2,\tau^2)}{\partial \tau^2} \frac{\partial \tau^2(h^2)}{\partial h^2_m}\right\vert_{h^2, \tilde{\tau}_{(1)}^2(h^2)}\\
    &= \quad \left. \frac{\partial l_{Y_{(1)}}(h^2,\tau^2)}{\partial h_m^2}\right\vert_{h^2, \tilde{\tau}_{(1)}^2(h^2)} + 0,
\end{align*}
where the last equality is due to $\tilde{\tau}_{(1)}^2(h^2)$ satisfying the first order condition for an interior partial maximizer; these calculations can be formalized \citep{milgrom2002envelope}. Thus, updating $\tau^2$ and then updating $h^2$ using the gradient of the log-likelihood, is equivalent to updating $h^2$ using the gradient of the profile log-likelihood. Using \eqref{eq:sigma_derivs}, $\partial l_{Y_{(1)}}(h^2,\tau^2) / \partial h_m^2$ is
\begin{align} \label{eq:h_deriv}
     -\frac{1}{2}\tr\left[\left\{\Sigma^{-1}_{(11)}(\theta) - \Sigma_{(11)}^{-1}(\theta)Y_{(1)}Y_{(1)}^\T \Sigma^{-1}_{(11)}(\theta)\right\} (K^m_{(11)} - I_{n_{(1)}})\right].
\end{align}

Similar arguments apply to the problem of finding $\hat{\theta}_0$. Let 
\[
    \Sigma_{(0\mid1)}(\theta) = \Sigma_{(00)}(\theta) - \Sigma_{(01)}(\theta)\Sigma^{-1}_{(11)}(\theta)\Sigma_{(10)}(\theta).
\]
Note that $\Sigma_{(0\mid 1)}(\theta) = \tau^2 \Psi_{(0\mid 1)}(h^2)$, where $\Psi_{(0\mid 1)}$ is defined like $\Sigma_{(0\mid 1)}$ but replacing every $\Sigma$ by $\Psi$. Thus, omitting the $h^2$ argument for simplicity, we have the partial maximizer
\begin{align*}
    \tilde{\tau}_{(0\mid 1)}^2 &= \argmax_{\tau^2 > 0}l_{Y_{(0)} \mid Y_{(1)}}(h^2, \tau^2) \\
    & = n_{(0)}^{-1} (Y_{(0)} - \Psi_{(01)}\Psi_{(11)}^{-1} Y_{(1)})^\T \Psi_{(0\mid 1)}^{-1} (Y_{(0)} - \Psi_{(01)} \Psi_{(11)}^{-1} Y_{(1)}).
\end{align*}
To apply gradient-based methods to the profile, conditional log-likelihood $h^2\mapsto l_{Y_{(0)}\mid Y_{(1)}}\{h^2, \tau^2_{(0\mid 1)}(h^2)\}$, observe
\[
    l_{Y_{(0)}\mid Y_{(1)}}(\theta) =  l_{Y}(\theta) -  l_{Y_{(1)}}(\theta).
\]
Derivatives of the two terms on the right-hand side can be obtained as in \eqref{eq:sigma_derivs} and \eqref{eq:h_deriv}.

In general, the objective functions we have discussed are nonconvex, and standard algorithms can be computationally expensive due to matrix decompositions needed to deal with the inverses, scaling approximately as $n^3$. Additionally, even when the parameter is identifiable in the distribution for $Y$, it can be unidentifiable, or nearly so, in the conditional distribution of $Y_{(0)}$ given $Y_{(1)}$, as we will see examples of later.

The constraint $\sum_{m = 1}^M h_m^2 < 1$ is also nontrivial in general. An exception is the case where $M = 1$ and $K^1_{(11)}$ is singular, because in that case any $h^2_1 \leq 0$ would lead to a $\Sigma_{(11)}(\theta)$ that is not positive definite, and hence an undefined or vanishing likelihood. Thus, for that case, the log-determinant term in the multivariate normal log-likelihood acts as a barrier. For the other cases, implementation is made easier by the fact that Theorem \ref{thm:validTest} continues to hold if we ignore the constraint when finding $\hat{\theta}_0$ and $\hat{\theta}_1$. Specifically, we may replace $\hat{\theta}_{0}$ by $\check{\theta}_0 = \{\check{h}_{(0\mid 1)}^2, \tilde{\tau}^2_{(0\mid 1)}(\check{h}_{(0\mid 1)}^2)\}^\T$, where
\[
    \check{h}_{(0\mid 1)}^2 \in \argmax_{h^2 \in \R{M}} l_{Y_{(0)}\mid Y_{(1)}}\{h^2, \tilde{\tau}^2_{(0\mid 1)}(h^2)\}.
\]
As noted before Theorem \ref{thm:validTest}, validity is retained since $\Llh_{Y_{(0)}\mid Y_{(1)}}(\check{\theta}_0) \geq \Llh_{Y_{(0)}\mid Y_{(1)}}(\hat{\theta}_0)$. Similarly, we can replace $\hat{\theta}_1$ by $\check{\theta}_1 = \{\check{h}_{(1)}^2, \tilde{\tau}^2_{(1)}(\check{h}^2)\}^\T$, where
\[
    \check{h}_{(1)}^2 \in \argmax_{h^2 \in \R{M}} l_{Y_{(1)}}\{h^2, \tilde{\tau}^2_{(1)}(h^2)\}.
\]
Validity is retained since $\check{\theta}_1$ is a function of $Y_{(1)}$ only. Both $\check{\theta}_0$ and $\check{\theta}_1$ can be obtained using unconstrained optimization.

In the following section, we discuss special cases of interest where computation can be made more efficient by using additional structure. 

\section{Testing Variance Components}\label{sec:Computationalimprovements}
%\section{Implementation}\label{sec:Implementation}
\subsection{Boundary Points}\label{sec:testBoundary}

Suppose we wish to test whether all but one variance component are zero; without loss of generality, $\sigma^2_M$ can be non-zero under the null hypothesis. Equivalently, $h^2_M$ can be non-zero under the null hypothesis. Thus, in the parameterization given by \eqref{eq:varcomp_sigma}, $\Theta_0$ is the set of $\theta = (h_1^2, \dots, h_M^2, \tau^2)^\T$ such that $h_1^2 = \cdots = h_{M - 1}^2 = 0$, $h_M^2 < 1$, and $\tau^2 > 0$. This setting is perhaps most natural when $M = 2$, in which case it corresponds to testing one variance component with the other unconstrained. Even that special case is challenging for existing methods when the unconstrained parameter is near the boundary.

The structure of $\Theta_0$ enables substantial computational gains compared to a naive implementation. In particular, let $K_M = O\Lambda O^\T$ by eigendecomposition and note that for any $\theta \in \Theta_0$,
\[
    \Sigma(\theta) = \tau^2\{h_M^2 K_M + (1 - h_M^2) I_n\} =  \tau^2 O \{h_M^2 \Lambda + (1 - h_M^2) I_n\} O^\T.
\]
Consequently, upon replacing $Y$ and $K_m$ by $O^\T Y$ and $O^\T K_m O$, respectively, $m \in \{1, \dots, M\}$, we may assume without loss of generality that $K_M = \Lambda$. With this assumption, $\Sigma(\theta)$ is diagonal for $\theta \in \Theta_0$, which simplifies computation of $\hat{\theta}_0$.

Specifically, when $K_M$ is diagonal, $\Sigma_{(01)}(\theta)$ is a matrix of zeros for $\theta \in \Theta_0$. Thus, for such $\theta$, 
$l_{Y_{(0)} \mid Y_{(1)}}(\theta) =  l_{Y_{(0)}}(\theta)$, which equals
\begin{align*}
    -\frac{1}{2}\sum_{k:Y_k\in Y_{(0)}}\left\{\log(\tau^2) + \log(h^2_M\lambda_k+ 1- h^2_M) \frac{Y_k^2}{\tau^2(h^2_M\lambda_k+1-h^2_M)}\right\},
\end{align*}
where $\lambda_k$ is the $k$th element of $\Lambda$ and, with a slight abuse of notation, $Y_k \in Y_{(0)}$ means the $k$th element of $Y$ is an element of $Y_{(0)}$. Thus, finding $\hat{\theta}_0$ reduces to a one-dimensional optimization problem over the interval $[0, 1)$ with an easy-to-compute derivative. This problem is substantially simpler than maximizing \eqref{eq:lY0|1_whole} in general, without diagonalization.

\subsection{Shared Eigenvectors}\label{sec:orthK}
%$\Sigma = \sigma^2(h_1^2K_1 + \cdots + h_M^2K_M + (1-h_1^2-\cdots-h_M^2)I_n)$, 
%$\Sigma = \sigma^2_1K_1 + \cdots + \sigma^2_MK_M + \sigma^2_eI_n$
%$\Sigma = \sigma^2(h_1^2\Lambda_1 + \cdots + h_M^2\Lambda_M + (1-h_1^2-\cdots-h_M^2)I_n)$
In some settings the eigenvectors of $K_m$ can be chosen not to depend on $m$; that is, one can find an orthogonal $O$ such that
\begin{equation}\label{eq:shared_eig}
         O^\T K_m O = \Lambda_m, ~~ m \in \{1, \dots, M\},
\end{equation}
where $\Lambda_m$ is a diagonal matrix with the eigenvalues of $K_m$ on the diagonal.

One such example is when the $K_m$ model variation in orthogonal directions, or more formally, $K_m K_\ell = 0$ for every $m\neq \ell$. Equivalently, by symmetry, every column of $K_m$ is orthogonal to every column of $K_\ell$. This claim follows from well-known facts about commuting, symmetric matrices; for completeness we give a more direct and constructive proof in the Appendix.

\begin{theorem}\label{thm:CoincideEigen}
    If symmetric matrices $K_1, \dots, K_M$ satisfy $K_m K_\ell=0$ for every $m\neq \ell$, then they satisfy \eqref{eq:shared_eig}.
\end{theorem}

Assuming \eqref{eq:shared_eig}, $O^\T Y$ has the multivariate normal distribution in \eqref{eq:mult_varcomp}, but with $\Lambda_m$ in place of $K_m$, $m \in \{1, \dots, M\}$. Thus, replacing $Y$ by $O^\T Y$ if needed, we may assume every $K_m = \Lambda_m$ is diagonal. We make this assumption for the remainder of the section. To set us up for a motivating example in the next section, we use the parameterization with $\sigma^2 = (\sigma_1^2, \dots, \sigma_{M + 1}^2)^\T \in \Omega =  [0, \infty)^M \times (0, \infty)$ and with some abuse of notation write $\Sigma(\sigma^2)$ for the covariance matrix of the multivariate normal distribution in \eqref{eq:mult_varcomp}.

Since $\Sigma(\sigma^2)$ is diagonal for every $\sigma^2 \in \Omega$, $l_{Y_{(i)}}(\sigma^2) = $ is, for  $i\in \{0, 1\}$,
\begin{equation} \label{eq:ll_diag_Lam}
   -\frac{1}{2}\sum_{k: Y_k \in Y_{(i)}}\left\{\log\left(\sum_{m = 1}^M \sigma_{m}^2 \lambda_{mk} + \sigma^2_{M + 1} \right) + \frac{Y_k^2}{\sum_{m = 1}^M \sigma_{m}^2 \lambda_{mk} + \sigma^2_{M + 1}} \right\},
\end{equation}
where $\lambda_{mk}$ is the $k$th diagonal element of $\Lambda_m$, $m \in \{1, \dots, M\}$. Moreover, $Y_{(0)}$ and $Y_{(1)}$ are independent, and hence $l_{Y_{(0)}\mid Y_{(1)}} = l_{Y_{(0)}}$, which simplifies finding $\hat{\theta}_0$ compared to the general case. The derivatives $\partial l_{Y_{(i)}}(\sigma^2) / \partial \sigma^2_m$ needed for gradient-based methods are, for  $i\in \{0, 1\}$,  
\[
    = -\frac{1}{2}\sum_{k: Y_k \in Y_{(i)}}\left\{\frac{\lambda_{mk}}{\sum_{r = 1}^M \sigma_{r}^2 \lambda_{rk} + \sigma^2_{M + 1} } - \frac{\lambda_{mk}Y_k^2}{\left(\sum_{r = 1}^M \sigma_{r}^2 \lambda_{rk} + \sigma^2_{M + 1}\right)^2} \right\},
\]
where $m \in \{1, \dots, M + 1\}$ and $\lambda_{(M+1)k} = 1$ for all $k$.

When $K_m K_\ell = 0$ further simplifications are possible: not only may we assume $K_m = \Lambda_m$, but after doing so it holds that for each $k$, there is at most one $m \in \{1, \dots, M\}$ for which $\lambda_{mk} \neq 0$. Let $\lambda_{(k)}$ be that $\lambda_{mk}$, with $\lambda_{(k)} = 0$ if no such $m$ exists, and let $\sigma^2_{(k)}$ be the corresponding $\sigma^2_m$. Then, for $i\in \{0, 1\}$
\[
        l_{Y_{(i)}}(\sigma^2) = -\frac{1}{2}\sum_{k: Y_k \in Y_{(i)}}\left\{\log\left(\sigma^2_{(k)}\lambda_{(k)} + \sigma^2_{M + 1} \right) + \frac{Y_k^2}{\sigma^2_{(k)}\lambda_{(k)} + \sigma^2_{M + 1}} \right\},
\]
and similarly for its derivatives.

\subsection{Crossed Random Effects}\label{sec:crossrandom}
Crossed random effects are known to be challenging both for theory and computation \citep{jiang2013subset,ekvall2020consistent,papaspiliopoulos2020scalable,ghosh2022scalable,lyu2024increasing,jiang2024precise, jiang2025asymptotic, ekvall2025uniform}. However, more efficient computing is possible by using a connection to the setting with shared eigenvectors, in the sense of \eqref{eq:shared_eig}. To introduce the setting, suppose momentarily that responses are naturally organized as a matrix $(Y_{ij})$ with $n_1$ rows and $n_2$ columns, with observations in the same row or column potentially dependent. A simple model with two crossed random effects is
\[
    Y_{ij} = U_{1i} + U_{2j} + E_{ij},
\]
where $U_{1i} \sim \rN(0, \sigma_1^2)$, $U_{2j} \sim \rN(0, \sigma^2_2)$, and $E_{ij} \sim \rN(0, \sigma^2_3)$, independently for all $i \in \{1, \dots, n_1\}$ and $j \in \{1, \dots, n_2\}$. The $U_{1i}$ can be interpreted as row effects and the $U_{2j}$ as column effects. 

More generally, suppose there are $M$ crossed random effects, with $n_m$ observations along the $m$th dimension, $m \in \{1, \dots, M\}$. Define the index set
\[
    \mathcal{J} = \{ (j_1, j_2, \dots, j_M) : j_m \in \{1, 2, \dots, n_m\} , m \in \{1, 2, \dots, M\} \},
\]
and suppose that for ${(j)} \in \mathcal{J}$,
\begin{equation}
    Y_{{(j)}} = U_{1j_1} + U_{2j_2} + \cdots+ U_{Mj_M} + E_{{(j)}},
\end{equation}
where $U_{mj_m} \sim \rN(0, \sigma^2_m)$ and $E_{{(j)}} \sim \rN(0, \sigma^2_{M + 1})$ are independent for all $m\in \{1, 2, \dots, M\}$ and $j_m\in \{1, 2, \dots, n_m\}$. Following \citet{ekvall2025uniform}, we can write this model as $Y = ZU + E$ by letting 
\[
    Z = (Z_1, \dots, Z_M), ~~ Z_m=1_{n_1}\otimes\cdots\otimes 1_{n_{m-1}} \otimes I_{n_m}\otimes 1_{n_{m+1}} \otimes\cdots\otimes1_{n_M},
\]
$m\in \{1, \dots, M\}$. Accordingly,
\[
    U = (U_1^\T, \dots, U_M^\T)^\T, ~~ U_m \sim \rN(0, \sigma^2_m I_{n_m}), ~~ m\in \{1, \dots, M\}.
\]
Now, to explore the connection to \eqref{eq:shared_eig}, define the projection matrices
$P_m=1_{n_m}1^T_{n_m}/n_m$, and $R_m^I = P_1\otimes\cdots\otimes P_{m-1}\otimes I_{n_m}\otimes P_{m+1}\otimes\cdots\otimes P_M$.  Let also $w_m = \prod_{k \neq m}n_k$. Then the covariance matrix of $Y$ is 
\begin{align*}
    \Sigma(\sigma^2) &= Z\ \cov_{\sigma^2}(U) Z^{\T} + \sigma^2_{M + 1}I_n \\
           &=\sum_{m=1}^M \sigma_m^2w_mR_m^I+\sigma_{M + 1}^2I_n.
\end{align*}
Define $Q_m = I_{n_m}-P_m$, $R_m^Q=P_1\otimes\cdots\otimes P_{m-1}\otimes Q_m\otimes P_{m+1}\otimes\cdots\otimes P_M$, and $R^P=P_1 \otimes\cdots\otimes P_M$. Then $R_m^I = R^P+R_m^Q$ and, consequently,
\begin{align*}
    \Sigma(\sigma^2) &= \sum_{m=1}^M\sigma_m^2w_mR_m^Q + \left(\sum_{m=1}^M\sigma_m^2w_m\right)R^P+\sigma_{M + 1}^2I_n.
           % &= \sum_{m=1}^M\sigma_m^2K_m + \left(\sum_{m=1}^M\sigma_m^2w_m\right)K_{M+1} + \sigma_e^2I_n
\end{align*}
Now note, for any $m \neq \ell$, by properties of Kronecker products, $P_m Q_m = 0$ and $R^P R_m^Q = R_m^Q R_\ell^Q = 0$. Thus, by Theorem \ref{thm:CoincideEigen}, there is an orthogonal $O$ such that $O^\T R_m^Q O = D^Q_m$ and $O^\T R^P O = D^P$, for diagonal $D_m^Q$, $m \in \{1, \dots, M\}$, and $D^P$. Thus, upon replacing $Y$ by $O^\T Y$, we may assume $\Sigma(\sigma^2)$ is diagonal. Specifically,
\begin{align*}
    \Sigma(\sigma^2) &= \sum_{m = 1}^M \sigma_m^2 w_m(D_m^Q + D^P) + \sigma_{M + 1}^{2}I_n \\
    &=  \sum_{m = 1}^M \sigma_m^2 \Lambda_m + \sigma_{M + 1}^{2}I_n,
\end{align*}
where $\Lambda_m = w_m(D_m^Q + D^P)$. Thus, inference can be based on \eqref{eq:ll_diag_Lam}. Note, however, that $\Lambda_m \Lambda_\ell = w_m w_\ell D^P \neq 0$ for $m\neq \ell$, so the further simplifications discussed following \eqref{eq:ll_diag_Lam} are not applicable.

The columns in the matrix $O$ can be computed relatively cheaply using that eigenvectors of $R_m^Q$, $m \in \{1, \dots, M\}$, are Kronecker products of eigenvectors of the matrices $P_1, \dots, P_{m-1}$, $Q_m$, $P_{m + 1}, \dots, P_M$. Specifically, there are $n_m - 1$ orthonormal eigenvectors corresponding to the eigenvalue one. Each of these is in the form $1_{m_-} \otimes v \otimes 1_{m_+} / \sqrt{w_m}$, where $v$ is an eigenvector of $Q_m$ corresponding to the eigenvalue one, and $1_{m_-}$ and $1_{m_+}$ are vectors of ones of lengths $m_- = \sum_{k = 1}^{m-1}n_k$ and $m_+ = \sum_{k = m + 1}^M n_k$, respectively. Additionally, one column of $O$ can be the eigenvector $1_{n}/\sqrt{n}$ of $R^P$. Thus, we have $1 + \sum_{m = 1}^M(n_m - 1)$ columns of $O$; the remaining can be obtained by completing the orthonormal basis in any way.

\subsection{Approximate Diagonalization}\label{sec:approxdiag}
In some settings \eqref{eq:shared_eig} may not hold exactly but approximately. That is, for an orthogonal $O$,  $\Vert \Lambda_m - O^\T K_m O\Vert$ is small for every $m$, and hence, intuitively, $O^\T Y$ has a covariance matrix that is close to diagonal. Similarly, $\Vert K_m K_\ell\Vert$ can be small for every $m \neq \ell$. In either case, it may be useful to replace $K_m$ by an approximation $\tilde{K}_m$, $m \in \{1, \dots, M\}$, such that the $\tilde{K}_m$ are jointly diagonalizable as in \eqref{eq:shared_eig}. To formalize, let us consider an example that we will examine in simulations.

Suppose $M = 2$ for simplicity, and that the $q_m < n$ greatest eigenvalues of $K_m$ are much larger than the trailing $n - q_m$, $m \in \{1, \dots, M\}$. Suppose also $K_1 = \Lambda_1$ is diagonal, with the diagonal elements of $\Lambda_1$ sorted in decreasing order. Suppose also that $K_2 = O_2 \Lambda_2 O_2^\T$, with the eigenvalues of $\Lambda_2$ sorted in decreasing order, and $O_2 = \bdiag(I_{q_2}, O_{(2)})$ for an orthogonal $O_{(2)}$ that is not the identity matrix. For simplicity we assume that the diagonal elements of each  are distinct. More specifically, the $q_m$ first elements of $\Lambda_m$ are evenly spaced on, say, $(a_2, a_3)$, and for some constant $c > 0$, the trailing $n - q_m$ eigenvalues are evenly spaced numbers on $(0, a_1)$ divided by a constant $c \geq 1$. The larger $c$ is, the better $K_m$ is approximated by the $\tilde{K}_m$ that sets small eigenvalues to zero. Notably, for those $\tilde{K}_m$, \eqref{eq:shared_eig} holds with $O = I_n$.

\section{Simulations}
\begin{figure}
    \centering
    \includegraphics[width=1\linewidth]{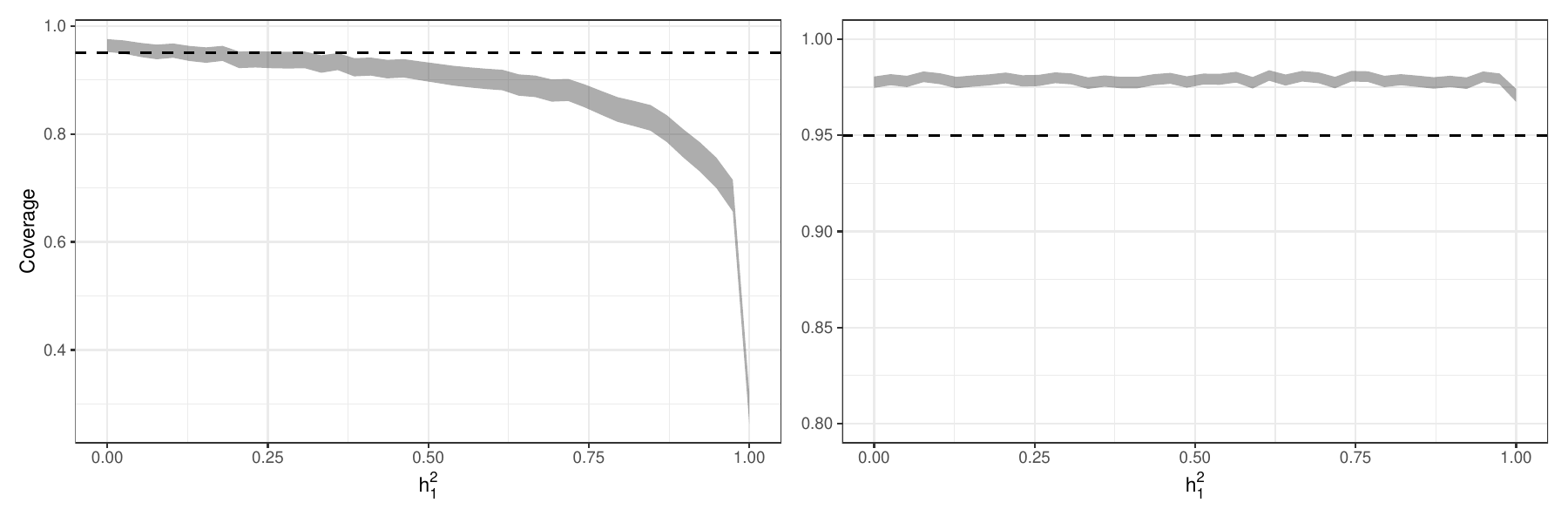}
    \caption{Coverage probabilities of a score-based confidence interval (left) and a randomized split LRT interval (right). The intervals are for $h_1^2$, with true values indicated on the horizontal axis. The true value of the nuisance parameter $h_2^2=0$. The dashed line is the nominal level 95\%. The Estimates are based on 10,000 replications.
    The shaded regions are 95\% confidence bands.}
    
    \label{fig:coveragechp3}
\end{figure}

Fig.\ref{fig:coveragechp3} shows, in a setting with $M = 2$ variance components, Monte Carlo estimates of the coverage probabilities for the score-based confidence interval discussed in the introduction and the proposed randomized split LRT confidence interval. The confidence intervals are for $h_1^2$, with $h^2_2$ a nuisance parameter. In the simulations, the matrices $K_1$ and $K_2$ were set to diagonal matrices with the eigenvalues of autoregressive correlation matrices with correlation parameters 0.95 and 0.5, respectively. That is, $K_1$ was diagonal with the eigenvalues of the $n\times n$ matrix $(0.95^{\vert i - j\vert})$, and similarly for $K_2$. We set $n = 300$, $h_2^2 = 0$, and $\tau^2 = 1$. The value of $h_1^2$ is on the horizontal axis. Monte Carlo estimates, and the corresponding confidence bands, are based on 10,000 replications.

Fig.~\ref{fig:coveragechp3} indicates that, when the nuisance parameter $h_2^2$ is near one, the score-based confidence interval for $h_1^2$ is invalid. The distortion is substantial, with coverage as low as 0.88 for extreme parameter values. By contrast, the proposed confidence interval, while conservative, is everywhere valid, as guaranteed by theory. The actual coverage probability of the proposed interval is around 0.975 regardless of the value of the nuisance parameter. For some values of the nuisance parameter this is substantially higher than the coverage probability of the score-based interval, but when $h_2^2$ is near zero the two intervals have similar coverage probabilities.

\begin{figure}
    \centering
    \includegraphics[width=1\linewidth]{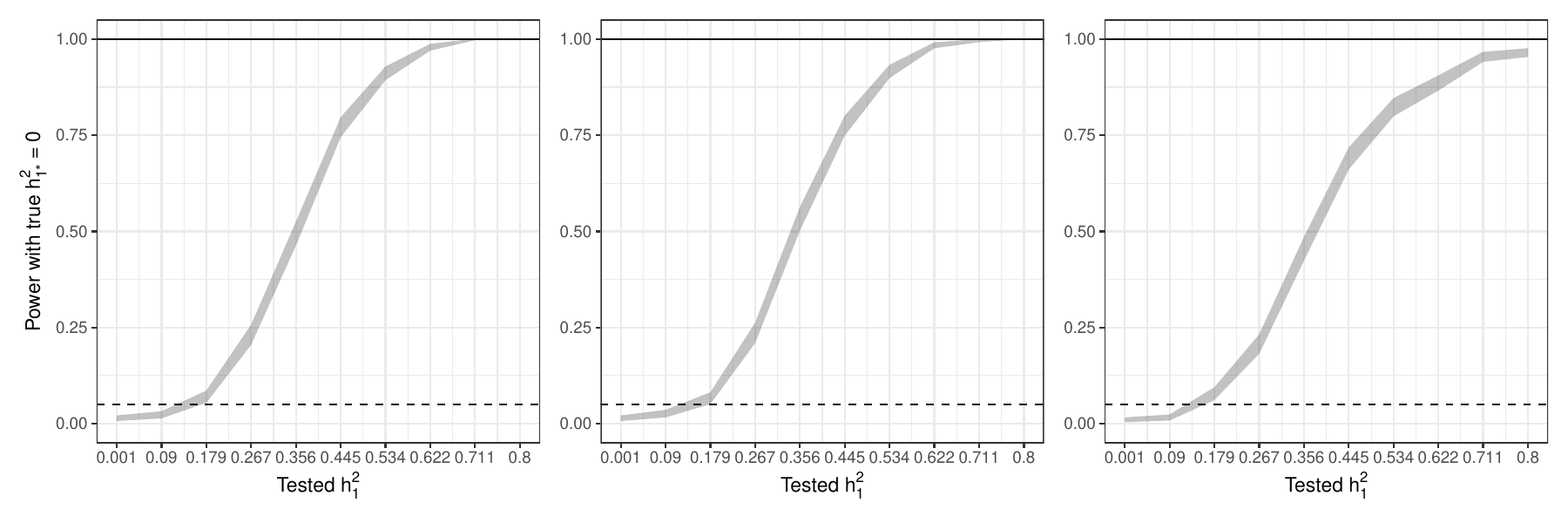}
    \caption{Rejection probabilities for split LRT with correctly specified model (left), approximated covariance matrices (middle), and unconstrained estimates (right). The true $h^2 = (0, 0.2)^\tsp$ and $\tau^2 = 1$. The dashed line is the nominal 5\% size. Estimates based on 1,000 replications. The shaded regions are 95\% confidence bands.}
    
    \label{fig:powerchp3}
\end{figure}

Fig.~\ref{fig:powerchp3} shows Mote Carlo estimates of rejection probabilities for the split likelihood ratio test in three different scenarios, all with $M = 2$ variance components. For all three scenarios, $K_1$ and $K_2$ were constructed as described in Sec.~\ref{sec:approxdiag}, with $a_1 = a_2 = 5$, $a_3 = 10$, and $c = 100$. Thus, the $q_m$th eigenvalue of $K_m$ in decreasing order is $5$ while the $(q_m + 1)$th is $5/100$. To ensure a non-identity $O_{(2)}$, we drew it, before starting the simulations, uniformly on the Stiefel manifold as the left singular vectors of a $(n - q_2) \times (n - q_2)$ matrix with independent standard normal entries. We set $q_1 = 100 \neq q_2 = 120$ to ensure identifiability. The true $h_1^2$, $h_2^2$, and $\tau^2$ were, respectively, 0, 0, and 1; the null hypothesis value of $h_1^2$ is on the horizontal axis.

The left plot in Fig.~\ref{fig:powerchp3} shows the proposed test is conservative--the rejection probability at $h_1^2 = 0$ is below the nominal level. As the null hypothesis value moves away from the truth, the rejection probability increases monotonically, as expected.

In the middle plot of Fig.~\ref{fig:powerchp3}, the proposed test is implemented with approximations $\tilde{K}_1$ and $\tilde{K}_2$ in place of the true $K_1$ and $K_2$ used to generate the data. Specifically, as discussed in Sec.~\ref{sec:approxdiag}, $\tilde{K}_m$ sets the $n - q_m$ smallest eigenvalues to zero, which leads to jointly diagonalizable $\tilde{K}_1$ and $\tilde{K}_2$. Thus, the test is much faster to implement than when using the true $K_1$ and $K_2$. Nevertheless, both size and power appear to be only minimally affected; the rejection probability curve is similar to that in the left plot, for which the true $K_1$ and $K_2$ were used.

In the right plot of Fig.~\ref{fig:powerchp3}, the true $K_1$ and $K_2$ are used, but $\hat{\theta}_0$ and $\hat{\theta}_1$ are replaced by, respectively, the estimators $\check{\theta}_0$ and $\check{\theta}_1$ which ignore the constraints on $h^2$. Because these estimators are unconstrained, they are simpler to compute using off-the-shelf solvers; see Sec.~\ref{sec:app_varcomp}. The rejection probability curve shows the test retains validity, as guaranteed by the arguments given before Theorem \ref{thm:validTest}. However, the power of the test is clearly lower than that of using constrained estimators. This is intuitive as the unconstrained estimators effectively ignore information about the parameter $h^2$.

% Fig.~\ref{fig:power_changeofc_chp3} shows rejection probability curves for experiments similar to those underlying the middle plot in Fig.~\ref{fig:powerchp3}, for different values of the constant $c$ and for true $h^2 = (0.4, 0.5)^\tsp$. Here we set $a_2=10$ and $a_3 = 20$ for $K_1$ to further amplify the differences between the two covariance matrices. Specifically, from left to right, $c = 0.1, 1, 10$. Recall, the larger $c$ is, the better we expect $\tilde{K}_1$ and $\tilde{K}_2$ to approximate $K_1$ and $K_2$, respectively. When $c = 0.1$, the eigenvalues set to zero are no longer the smallest ones, and that leads to an unreliable test with size close to 0.8 (left plot). When $c = 1$ the test has approximately correct size, but power is lower than size for many values of the tested parameter. When $c = 10$, the test is valid and has some power if the tested parameter is sufficiently far from the truth. A likely reason for the stark difference between the results in Fig.~\ref{fig:power_changeofc_chp3} and the middle plot in Fig.~\ref{fig:powerchp3} is that, in Fig.~\ref{fig:powerchp3}, the true $h_1^2$ is zero, and $h_2^2$ is relatively small, so the distribution of the data does not depend on $K_1$ and depends to a smaller extent on $K_2$.

\begin{figure}
    \centering
    \includegraphics[width=1\linewidth]{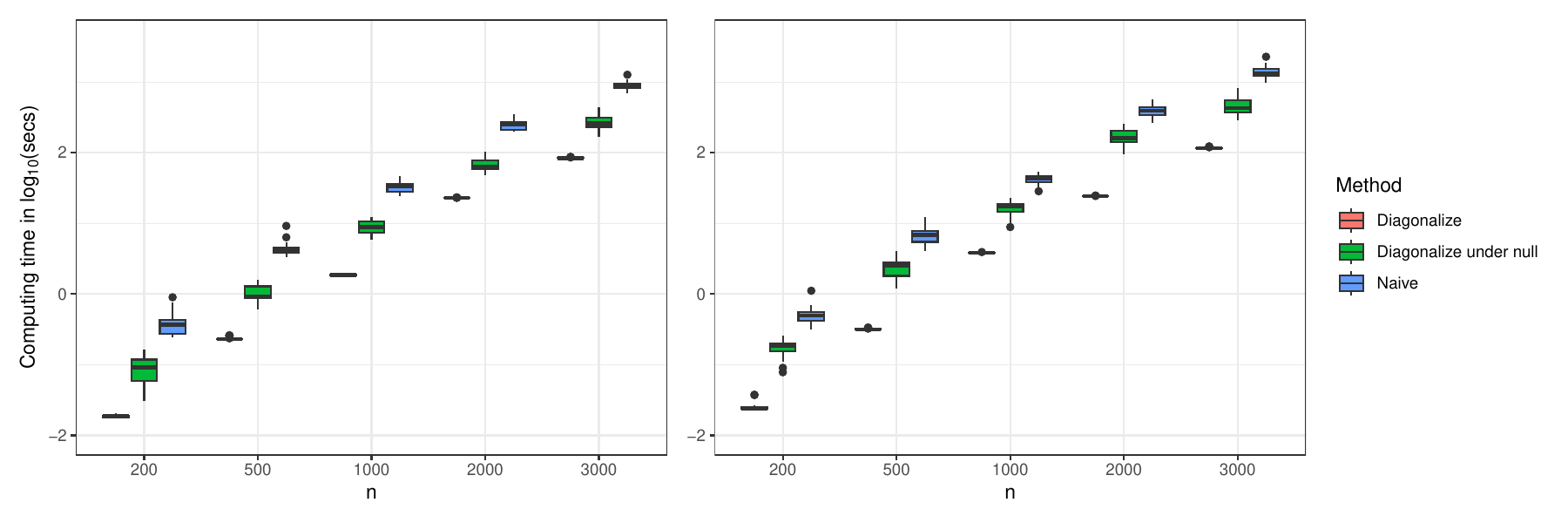}
    \caption{Computing times for three implementations of the split likelihood ratio test with $M = 2$ (left) and $M = 3$ (right) components.}

    \label{fig:sim}
\end{figure}

Fig.\ref{fig:sim} shows computing times for different implementations of the proposed split LRT statistic. We consider a setting where the $K_m$ are jointly diagonalizable as in \eqref{eq:shared_eig}. Specifically, we generated $K_m$, $m \in \{1, \dots, M\}$ as follows. First, we eigendecomposed an $n\times n$ autoregressive covariance matrix $A = (A_{ij}) = (0.5^{\vert i - j\vert}) = O \Lambda O^\tsp$. Then, $\Lambda_m$ was created by setting $\lfloor n / (M + 1) \rfloor$ of its diagonal entries equal to the corresponding entries of $\Lambda$, and the remaining entries to zero. The indices for the non-zero entries for each $\Lambda_m$ were randomly sampled in such a way that they were different for each $\Lambda_m$, so that $\Lambda_m \Lambda_\ell = 0$ for $\ell\neq m$, and $\sum_{m = 1}^M \Lambda_m$ equals $\Lambda$ with $n - M \lfloor n / (M + 1)\rfloor$ entries set to zero. Finally, we set $K_m = O \Lambda_m O^\tsp$, $m \in \{1, \dots, M\}$, so $K_m K_\ell = 0$ for $\ell \neq m$. When there are $M = 2$ components (left plot), data are generated with $h^2 = (0, 0.2)^\tsp$ and the null hypothesis is $h_1^2 = 0$. When $M = 3$, data are generated with $h^2 = (0, 0, 0.2)^\tsp$ and the null hypothesis is $h_1^2 = h^2_2 = 0$.

The three methods in Fig.~\ref{fig:sim} include a naive method that implements the split likelihood ratio test as described in Sec.~\ref{sec:app_varcomp}, without using the fact that the $K_m$ can be jointly diagonalized. The second considered method uses this fact only when calculating $\hat{\theta}_0$, emulating a setting where diagonalization is possible under the null hypothesis, but not in general, as in Sec.~\ref{sec:testBoundary}. The third method uses joint diagonalization both when computing $\hat{\theta}_0$ and $\hat{\theta}_1$. For small sample sizes, the methods are all fast and hence no large differences are seen in computing times. In contrast, when $n$ is in the thousands, using diagonalization leads to substantially faster computing. For example, with $M = 2$ components and $n = 3000$, the naive method takes on average about 900s, while the method that uses diagonalization under the null only takes about 270s on average, and the method which fully uses diagonalization takes about 80s on average. The plots are consistent with the facts that evaluating the likelihood and its derivatives takes $O(n)$ operations when using diagonalization and up to $O(n^3)$ operations otherwise.

The simulation results, along with those of the data example in the next section, can be reproduced using code at \url{https://github.com/koekvall/univ_vc_suppl}.
 
% \begin{figure}
%     \centering
%     \includegraphics[width=1\linewidth]{Figures/power_variousc_0.4&0.5.pdf}
%     \caption{Rejection probabilities for split LRT using approximated covariance matrices with different constant $c$. The true $h^2 = (0.4, 0.5)^\tsp$ and $\tau^2 = 1$. The dashed line is the nominal 5\% size. Estimates based on 5,000 replications. The shaded regions are 95\% confidence bands.}
%     \label{fig:power_changeofc_chp3}
% \end{figure}

% \begin{figure}
%     \centering
%     \includegraphics[width=1\linewidth]{Images/power_variousc_0.1-1-10.pdf}
%     \captionof{figure}[]{Plot}
%     \label{fig:power_changeofc_chp3}
% \end{figure}

\section{Data Example}

To illustrate the proposed methods, we apply them to a well-known dataset \cite[Problem 6.18]{Hicks1999-jx}. The data consist of resistance measurements (in milliohms) obtained through a fully randomized design. Ten resistors and three operators were randomly selected. Each operator independently measured the resistance of each resistor twice, resulting in a total of 60 observations. Both resistor and operator effects are treated as random in the analysis. This can be motivated, for example, by thinking of the operators and resistors in the experiment as drawn from larger populations of potential operators and resistors, respectively. The random effects model correlations between measurement from the same operator, and measurements from the same resistor. Because every operator measures every resistor, the random effects are crossed. Specifically, suppose the $k$th measurement by operator $j$ on resistor $i$ satisfies
\[
    \tilde{Y}_{ijk} = \beta + U_{1i} + U_{2j} + E_{ijk},~~ (i, j, k) \in \{1, \dots, 10\}\times \{1, 2, 3\} \times \{1, 2\},
\]
where $\beta \in \R{}$ is the population mean, $U_{1i} \sim \rN(0, \sigma^2_1)$ is the resistor random effect, and $U_{2j} \sim \rN(0, \sigma^2_2)$ is the operator random effect. For simplicity, we estimate $\beta$ using the sample mean and apply our method to the centered responses $Y_{ijk} = \tilde{Y}_{ijk} - \sum_{i, j, k}\tilde{Y}_{ijk}/60$. That is, we fit the model
\begin{equation}
    Y \sim \rN(0, \sigma^2_1 I_{10} \otimes 1_21_2^\T \otimes 1_31_3^\T + \sigma^2_2 1_{10} 1_{10}^\T \otimes 1_21_2^\T \otimes I_{3}+\sigma^2_3 I_{60}),
\end{equation}
where $Y$ is obtained by stacking the $Y_{ijk}$.

Maximum likelihood estimates of $\sigma^2$, computed with the \texttt{lme4} package \citep{bates2015fitting}, are in Table \ref{tab:est_hicksturner}. The table also includes the corresponding estimates of $h^2$ and $\tau^2$ (``total variance"). The estimate of $\sigma_1^2$, the variance of the resistor random effect, is zero. The estimate of the operator random effect $\sigma^2_2$ is about 50, which leads to an estimate of $h_2^2$ of 0.554. That is, it is estimated that about 55\% of the total variability is due to the operator random effect.

\begin{table}[t!]
    \caption[Maximum likelihood estimates for the resistor data]{Maximum likelihood estimates of variance components and proportions of variability for the resistor data. Standard errors based on observed information are in parentheses.}\label{tab:est_hicksturner}
    \begin{tabularx}{6.5in}{XXXX}
      \hline
      &resistor&operator&error/total variance  \\
      \hline
      $\sigma^2$ & 0.00 (5.91) & 50.0 (42.5) & 40.3 (7.74) \\
      $h^2$ & 0.00 (0.0654) & 0.554 (0.217) & 90.3 (43.4) \\	
      \hline
    \end{tabularx}
\end{table}

Table \ref{tab:cis_hicksturner} shows confidence intervals for the random effect standard deviations $\sigma_1$ and $\sigma_2$; confidence intervals for the variances can be obtained by squaring the endpoints. The table includes three standard methods, two of which are included by default in \texttt{lme4}, namely profile likelihood and bootstrap-based intervals; and the proposed method based on the split LRT. The Wald interval is not available by default in {\tt lme4} since it is known to be unreliable, but we nevertheless compute it for comparison. All methods give confidence intervals for $\sigma_1$ that include zero. The confidence intervals for $\sigma_2$ are relatively large, likely due to the small number of operators (three). However, only one of the intervals for $\sigma_2$, the Wald interval, includes zero. For both parameters, the proposed interval is substantially wider than the other three. However, it is also the only one that is known to be valid.

\begin{figure}
    \centering
    \includegraphics[width=1\linewidth]{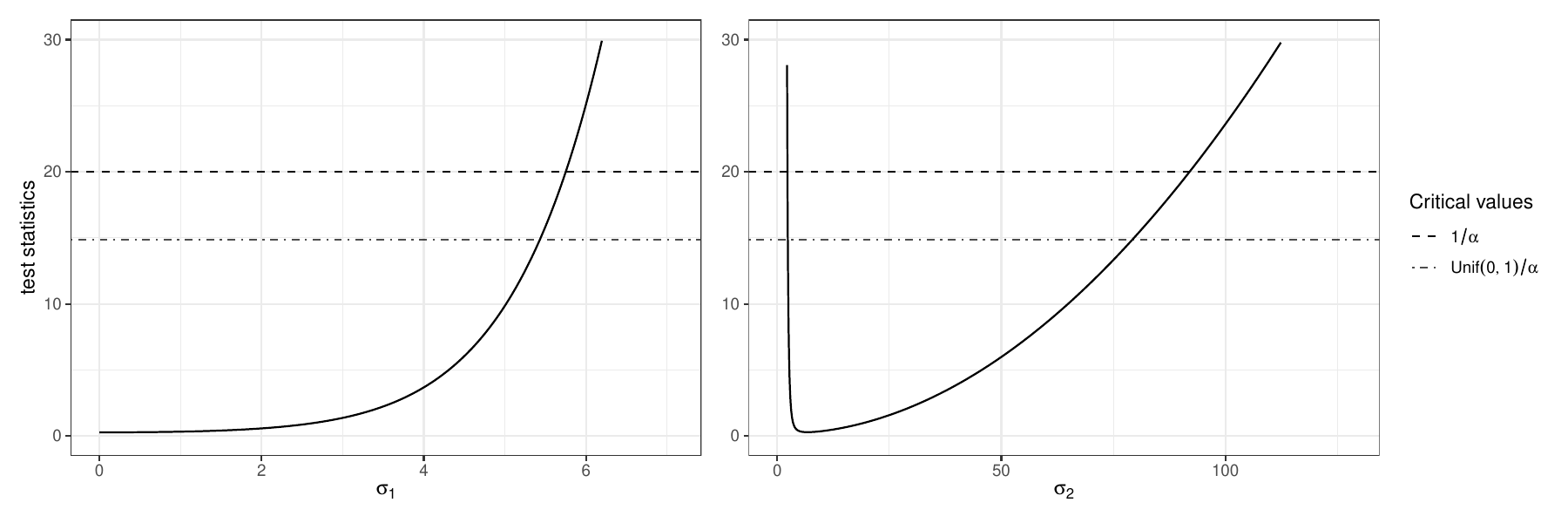}
    \caption{Graphs of test statistics for $\sigma^2_1$ (left) and $\sigma^2_2$ (right) with the resistor data. The dashed and dot-dash lines are, respectively, thresholds of split LRT ($1/\alpha = 20$) and randomized split LRT ($\rU(0,1)/\alpha) = 0.742 / 0.05$).}
    
    \label{fig:CI_hicksturner}
\end{figure}

To compute confidence intervals with the proposed method, given the small number of observations, we use a $k$-fold method that can reduce the variability introduced by random data splits \citep{wasserman2020universal}. We use $k=4$ folds to strike a balance between reducing randomness and retaining sufficient data within each fold. In each iteration, one fold is used to compute $\hat{\theta}_1$ and the three other folds are used to compute $\hat{\theta}_0$. Then the average of the four test-statistics are compared to the threshold $U / \alpha$ (c.f. \ref{eq:test_stat_slrt}). Figure \ref{fig:CI_hicksturner} shows graphs of the resulting test-statistics for a range of $\sigma_1$ and $\sigma_2$. The confidence intervals contain the points where the corresponding graphs are below the critical value, which is drawn as a horizontal line. The particular realization of $U$ was 0.742, so the critical value is $0.742 / 0.05 \approx 15$. For comparison, recall the non-randomized test has critical value $1 / \alpha = 20$.

 \begin{table}[t!]
    \caption[Confidence intervals for the resistor data]{Confidence intervals for square-root variance components for the resistor data.}\label{tab:cis_hicksturner}
    \begin{tabularx}{6.5in}{XXXXX}
      \hline
       & Wald & profile & bootstrap & split LRT \\
      \hline
      $\sigma_1$& (0.00, 3.40) & (0.00, 2.80) & (0.00, 2.72) & (0.00, 5.43)\\
      $\sigma_2$& (0.00, 11.5) & (3.55, 21.3) & (1.16, 12.3) & (2.38, 78.8) \\	
      \hline
    \end{tabularx}
\end{table}

We can find $p$-values for the null hypotheses $\sigma_m^2 = 0$, $m \in \{1, 2\}$, by finding the smallest $\alpha \in (0, 1)$ for which the tests would reject, with the convention that the $p$-value is one if no such $\alpha$ exists. For the randomized split LRT, the $p$-value for $\sigma_1 = 0$ is one while the $p$-value for $\sigma_2 = 0$ is zero. For comparison we used the \texttt{exactRLRT} function from the R package \texttt{RLRsim}, which provides likelihood ratio tests for boundary points based on simulation. The $p$-values for $\sigma_1 = 0$ and $\sigma_2 = 0$ are 0.452 and 0.00, respectively.

\begin{figure}
    \centering
    \includegraphics[width=1\linewidth]{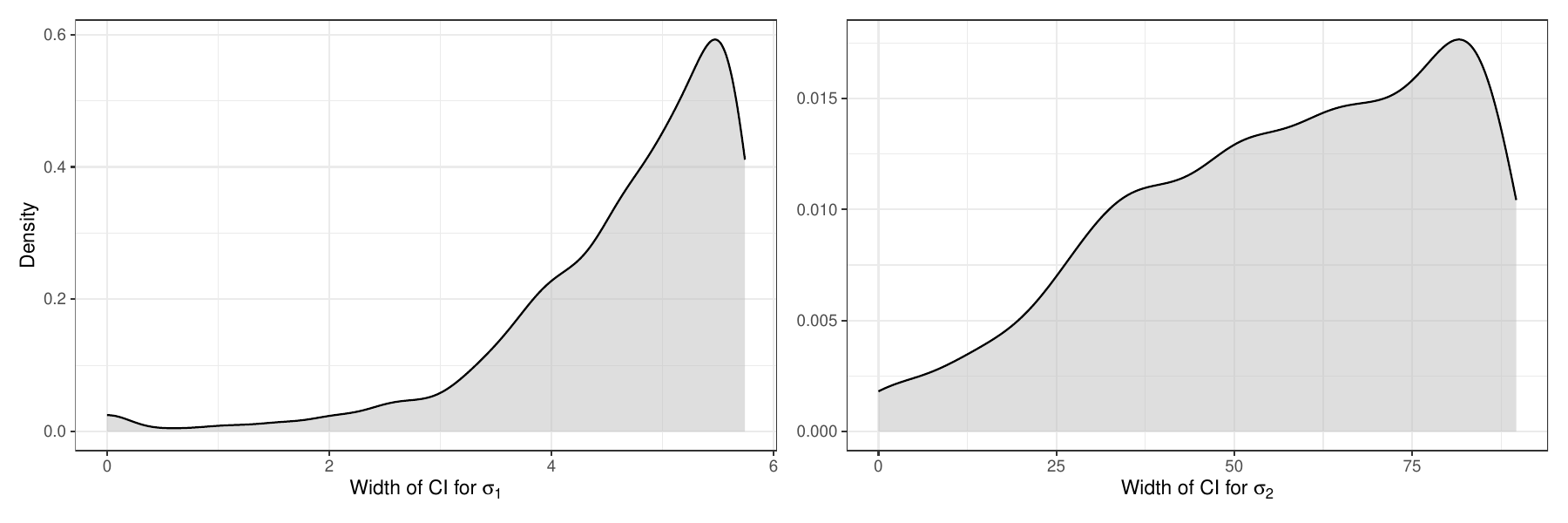}
    \caption{Densities for the distributions of widths of randomized split LRT confidence intervals for $\sigma_1$ (left) and $\sigma_2$ (right) with the resistor data. Based on drawing 1,000 random thresholds $\rU(0, 1)/\alpha$ and computing the width for each threshold.
    }
    
    \label{fig:hicksturner_CIlength}
\end{figure}

Because the critical value for the randomized test in general is different each time it is implemented, $p$-values and widths of confidence intervals also vary from implementation to implementation. In Figure \ref{fig:hicksturner_CIlength} we examine the distribution of the widths of the confidence intervals in this data example. Because the non-randomized test compares to $1/\alpha$ rather than $U/\alpha$, its width is the greatest value on the horizontal axis where the density has support, which corresponds to a realization $U = 1$. For example, the right plot in Fig.~\ref{fig:hicksturner_CIlength} indicates the non-randomized CI for $\sigma_2$ has width of about 90, as also seen in Fig.~\ref{fig:CI_hicksturner}. The distribution of the width of the randomized CI has a mode around 80, which happens to be about the length we observed in Table \ref{tab:cis_hicksturner}. Using Monte Carlo, we found that the randomized CI widths were on average $81.1\%$ and $63.7\%$ of the non-randomized ones, for $\sigma_1$ and $\sigma_2$ respectively. 

\begin{figure}
    \centering
    \includegraphics[width=1\linewidth]{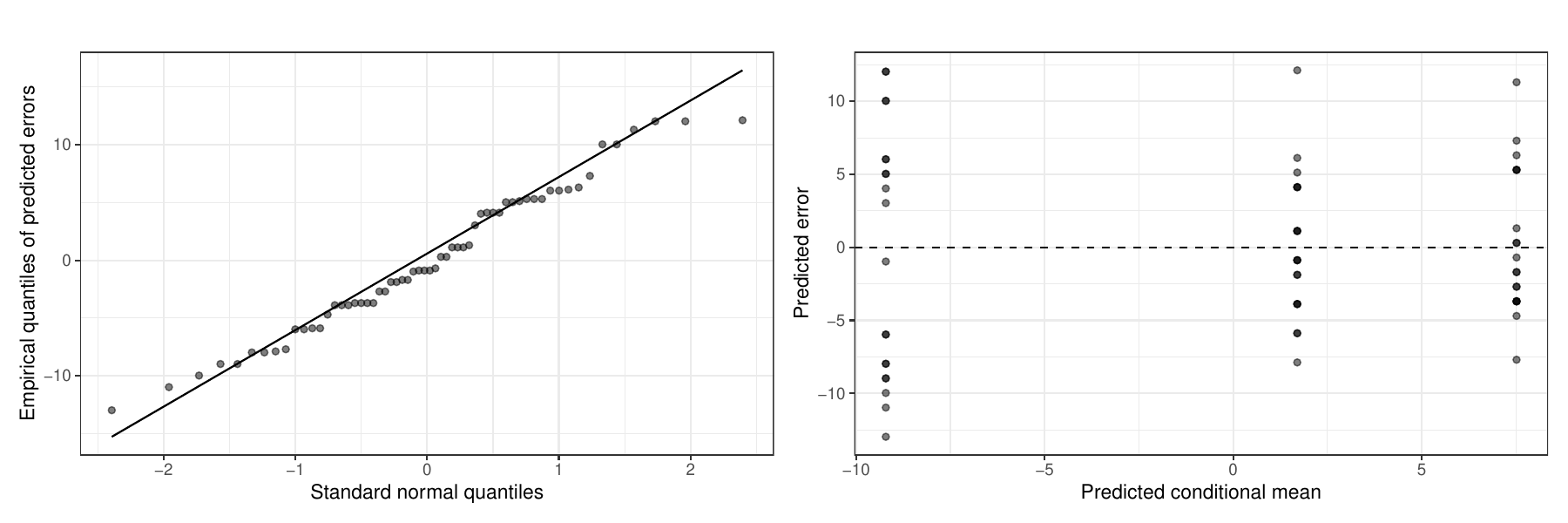}
    \caption{Diagnostic plots using predicted errors in the resistor data.}
    
    \label{fig:diagnostic_hicksturner}
\end{figure}

Finally, since the proofs of validity of the proposed methods require a correctly specified likelihood, we consider diagnostic plots. With $G =\bdiag(\hat{\sigma}^2_1 I_{10}, \hat{\sigma}^2_2I_{3})$, we can predict the random random effects with the best linear unbiased predictions (BLUPs) \citep{robinson1991blup}, i.e.,
\[
    \hat{U} = G Z^{\T}(ZGZ^{\T}+\hat{\sigma}^2_3 I_{60})^{-1}Y.
\]
Thus, we can predict $\E(Y\mid U) = ZU$ by $Z \hat{U}$, and $E = Y - ZU$ by $\hat{E} = Y - Z \hat{U}$. If the predictions are accurate and the model is correct, we expect $\hat{E}$ to be distributed approximately as $E\sim \rN(0, \sigma_3^2 I_n)$; the quantile-quantile plot in Fig.\ref{fig:diagnostic_hicksturner} supports this approximation. The right plot in Fig.~\ref{fig:diagnostic_hicksturner} shows $\hat{E}$ plotted against the predictions $Z\hat{U}$; there are only three levels because $\hat{\sigma}^2_1 = 0$. The variability in the $\hat{E}_i$ is perhaps slightly larger for the smallest value of $\hat{Y}_i$ compared to the other two, but overall the assumption that $E_i$ is independent of $Z U$, with  $E_i \sim \rN(0, \psi_3)$, appears serviceable.

\section{Conclusions}

The proposed methods lead to valid inference on variance components even in settings where existing methods fail. In particular, to the best of our knowledge, it is the first method to be uniformly valid in settings where heritability, or a proportion of variation more generally, is near unity. The main drawback of the proposed methods is that they are conservative, but that is not a problem if a test rejects or if the confidence interval is narrow enough to be useful. Whether a confidence interval is narrow enough to be useful has to be assessed on a case-by-case basis by practitioners.

In many settings of interest the eigenvectors of the covariance matrix of the response vector do not depend on the variance components. Then, the algorithms provided here lead to much faster computing than naive ones. Further improvements may be possible with more research. For example, it is unclear how different methods for splitting the data affect the methods. It would be of interest to understand, for example, whether diagonalization is best performed before or after splitting. Similarly, in settings with crossed random effects, it may be preferable to, instead of splitting uniformly at random, balance the randomization so that an equal number of levels of a given factor is present in both splits. Finally, there may be room for improvements in both size and power by using estimators other than maximum likelihood. We focused on maximum likelihood estimators because they are common and have good large sample properties, but there are certainly many settings where we expect better estimation is possible. For instance, penalized likelihood-based estimators can be preferable in settings with many parameters.

\bibliographystyle{apalike}
\bibliography{univ_vc}

\appendix 
\section{Technical details}

\begin{proof}[Proof of Theorem 1]
    The arguments are available in the literature \cite[Theorems 1.2 and 3, respectively]{ramdas2023randomized, wasserman2020universal}. Pick an arbitrary $\theta^* \in \Theta_0$. By Markov's inequality,
    \[
     \pr_{\theta^*}\left(T_n> 1 / \alpha\right) \leq   \alpha \E_{\theta^*}\left(T_n\right),
    \]
    so the first claim follows if $\E_{\theta^*}\left(T_n\right) \leq 1$.  To show the latter, note that since $\theta^* \in \Theta_0$ and $\hat{\theta}_0$ is a maximizer over that set,
    \[
         \E_{\theta^*}\left(T_n\right) = \E_{\theta^*}\left\{\frac{\Llh_{Y_{(0)} | Y_{(1)}}(\hat{\theta}_1)}{\Llh_{Y_{(0)} | Y_{(1)}}(\hat{\theta}_0)}\right\} \leq  \E_{\theta^*}\left\{\frac{\Llh_{Y_{(0)} | Y_{(1)}}(\hat{\theta}_1)}{\Llh_{Y_{(0)} | Y_{(1)}}(\theta^*)}\right\}.
    \]
    Thus, it suffices to show that the last expectation equals one. To that end, condition on $Y_{(1)}$ and note that, since 
    $\hat{\theta}_1$ is measurable with respect to the $\sigma$-algebra generated by $Y_{(1)}$, with probability one,
    \begin{align*}
    \E_{\theta^*}\left\{\frac{\Llh_{Y_{(0)} | Y_{(1)}}(\hat{\theta}_1)}{\Llh_{Y_{(0)} | Y_{(1)}}(\theta^*)}\bigg|Y_{(1)}\right\} & = \int \frac{f_{\hat{\theta}_1}(y_{(0)}|Y_{(1)})}{f_{\theta^*}(y_{(0)}|Y_{(1)})}f_{\theta^*}(y_{(0)}|Y_{(1)})dy_{(0)}\\%\cdots dy_{0\lfloor\frac{n}{2}\rfloor}\\
    & = \int f_{\hat{\theta}_1}(y_{(0)}|Y_{(1)})dy_{(0)}= 1. 
    \end{align*}
    Here, $f_\theta(y_{(0)}\mid Y_{(1)})$ denotes the density of the conditional distribution of $Y_{(0)}$ given $Y_{(1)}$ under $\theta$, evaluated at a fixed $y_{(0)}$ and random $Y_{(1)}$. Since the conditional expectation equals one, so does the unconditional one, and that proves validity of the split likelihood ratio test.

    The proof of validity of the randomized split likelihood ratio test is similar except for the first step. Specifically, instead of using Markov's inequality we note that by independence of $U$ and $T_n$,
    \begin{align*}
       &\pr_{\theta^*}\left(T_n>U / \alpha\right)  =\E_{\theta^*}\left\{\pr_{\theta^*}(U<\alpha T_n \mid T_n)\right\} =\E_{\theta^*}\left\{\min(\alpha T_n, 1)\right\} \\ &\leq   \alpha \E_{\theta^*}\left(T_n\right).
    \end{align*}
    Thus, validity again follows from $\E_{\theta^*}\left(T_n\right) \leq 1$.

    Finally, the claim about power is immediate from the fact that the event $\{\alpha T_n > 1\}$ is a subset of the event $\{\alpha T_n > U \}$ since $U$ has support $(0, 1)$.
\end{proof}

\begin{proof}[Proof of Theorem 2]
    First note that, by symmetry, each $K_m$ has an eigendecomposition $K_m = O_m \Lambda_m O_m^\T$, where $\Lambda_m = \diag(\lambda_{m1}, \dots, \lambda_{mn})$. Since the conclusion is obvious if $K_m = 0$ for all $m$, suppose not. Pick arbitrary $K_m$ and $K_\ell$, $m\neq \ell$, such that $\lambda_{\ell k} \neq 0$ is an element of $\Lambda_\ell$ and $o_{\ell k}$ the corresponding column of $O_\ell$. Thus, $K_\ell o_{\ell k} = \lambda_{\ell k} o_{\ell k}$. Left-multiplying by $K_m$ and using $K_m K_\ell = 0$ leads to
    \[
        0 = K_m K_\ell o_{\ell k} = \lambda_{\ell k} K_m o_{\ell k},
    \]
    so $o_{\ell k}$ is an eigenvector of $K_m$, corresponding to the eigenvalue zero. Thus, $o_{\ell k}$ is orthogonal to every column of $O_m$ corresponding to a nonzero eigenvalue of $K_m$. Take now every vector that, for some $m \in \{1, \dots, m\}$, is a column of $O_m$ corresponding to a nonzero eigenvalue of $K_m$, say
    \[
      \mathcal{O} = \{o_{m k}: \lambda_{mk} \neq 0 \text{ for some } m\in \{1, \dots, M\} \text{ and} ~k\in \{1, \dots, n\}\}.
    \]
  This set is orthonormal. Indeed, we already showed that any two such vectors from different $O_m$ are orthogonal, and if they are from the same $O_m$ they are orthogonal by construction. Because $\mathcal{O} \subseteq \R{n}$ is orthonormal, it contains at most $q \leq n$ vectors. Let these vectors be the leading $q$ columns of $O$ (in any order), and take the remaining $n - q$ vectors to be any orthonormal basis for the orthogonal complement of the span of $\mathcal{O}$. Note the last $n - q$ vectors are in the null space of every $K_m$, $m \in \{1, \dots, M\}$ by construction of $\mathcal{O}$. Now \eqref{eq:shared_eig} holds upon possibly reordering elements in $\Lambda_m$, $m \in \{1, \dots, M\}$.
\end{proof}
\end{document}